# Study on the efficiency droop in high-quality GaN material under high photoexcitation intensity

Peng Chen*, Zili Xie, Xiangqian Xiu, Dunjun Chen, Bin Liu, Hong Zhao, Yi Shi, Rong Zhang, and Youdou Zheng

*Jiangsu Provincial Key Laboratory of Advanced Photonic and Electronic Materials and School of Electronic Science and Engineering, Nanjing University, Jiangsu, Nanjing, 210093, China.*
*Authors to whom correspondence should be addressed: pchen@nju.edu.cn

## Abstract

III-V nitride semiconductors, represented by GaN, have attracted significant research attention. Driven by the growing interest in smart micro-displays, there is a strong desire to achieve enhanced light output from even smaller light-emitting diode (LED) chips. However, the most perplexing phenomenon and the most significant challenge in the study of emission properties under high-injection conditions in GaN has always been *efficiency droop* for decades, where LEDs exhibit a substantial loss in efficiency at high driving currents. In this paper, we present our study on the intrinsic emission properties of high-quality GaN material based on the density of states and the principles of momentum conservation. Our theoretical calculations reveal a momentum distribution mismatch between the non-equilibrium excess electrons and holes, which becomes more significant as the carrier concentration increases. Our excitation-dependent photoluminescence measurements conducted at 6 K exhibited a clear droop for all exciton recombinations, but droop-free for phonon-assisted recombination due to phonons compensating for the momentum mismatch. These findings indicate that the momentum distribution mismatch between the non-equilibrium excess electrons and holes is one of the intrinsic causes of the efficiency droop, which originates from the intrinsic band properties of GaN. These results suggest that proper active region design aimed at reducing this mismatch will contribute to the development of ultra-highly efficient lighting devices in the future.

## 1. Introduction

GaN-based III–V nitride semiconductors have been subjected to intense research over the past decades, yielding tremendous progress in the fabrication of light-emitting diodes (LEDs)[1–3] and laser diodes (LDs)[4, 5] and a wide range of applications in Solid State Lighting (SSL), lighting engineering, display, water treatment, sterilization of medical equipment and many other important applications that have greatly enhanced societal and human life. As a result, the 2014 Nobel Prize for Physics has been awarded to Isamu Akasaki, Hiroshi Amano and Shuji Nakamura for their achievement in GaN-based blue LEDs, and the GaN material is the core material for the LEDs. Although the LEDs have been proven to be more efficient than conventional incandescent light sources [6, 7], it is still very important to continuously improve LED lighting efficiency in smaller LED chips and under high injection condition for , the cost of light can be reduced further.

The crucial issue is to enhance the internal quantum efficiency (IQE), the ratio of the number of photons emitted per second to the number of carriers injected per second. But, researchers have encountered a big problem since 2001, which is called *efficiency droop*[8, 9], a significant efficiency decrease in the LED with the increase of the forward current to high current density, more than 70% loss in at high current density in some cases. The droop is particularly great damage to the applications of GaN-based LEDs. It is therefore imperative to understand and overcome the efficiency droop in GaN research.

GaN-based LEDs normally consist of multiple hetero-epitaxial layers including multiple-quantum-well (MQW), AlGaN electron-blocking layer, etc. The complex internal structure generated a lot of possibilities resulting in the efficiency droop, including polarization fields[10, 11], density-activated defect tunneling[12], asymmetry of carrier transportation[13, 14], and Auger processes[15-19]. etc. Up to now, the LED technical community does not have gotten recognized conclusion for the droop. The most fundamental reason still is unclear.

Here, we analyze the problem from a different aspect, i.e. from GaN material itself. In principle, the radiative recombination is a kind of intrinsic behavior between excess electrons and holes. So, it has become urgent to understand the problem from the material intrinsic process without any external influence. Because the intrinsic recombination behavior can be covered up by high defect density and sundry structure design in normal GaN-based LEDs, a high quality GaN material without any heterostructure is required to eliminate well-known factors, such as polarization, carrier transportation and so on. At the same time, an accurate measurement method is also required to avoid unwanted effects such as heat, asymmetrical transportation and so on. Here, photoluminescence (PL) method can provide a symmetrical injection for electrons and holes without carrier transportation concerns, and the most defect-related effects can also be avoided when the PL measurements are carried out at low temperature.

In this study, a single GaN layer grown on a freestanding GaN substrate (GaN/GaN) is used for PL measurements. The single GaN layer has a very low dislocation density without any heterostructure. By solving a 6×6 K·P Hamiltonian, we obtained the GaN energy band structure, the density of states (DOS), and furthermore obtained the distribution of nonequilibrium excess carriers in the valance band (VB) and conduction band (CB) in GaN. We adopted varied excitation PL measurements on the high quality GaN/GaN layer at 6 K. From all results, we investigated the efficiency droop based on GaN intrinsic band properties and conservation of energy/momentum for electron-hole recombination.

## 2. Theoretical calculations

By solving the 6×6 K·P Hamiltonian, GaN CB and VB are obtained (Supplementary Fig. S1). Similar to existing results, alone $k_z$ direction and $k_x$ direction, the energy bands are asymmetric. After calculation of the DOS of each band (Fig.1), the carrier occupation in the CB and VB can be obtained as shown in Fig.1B, in which thermal excitation is ignored at low temperature. We set the edge of CB or VB to 0. From Fig.1, it can be clearly seen that the DOS of heavy hole band (HH) is larger than that of light hole band (LH) but still in the same order of magnitude. So, the HH band dominates the hole distribution in relative low concentration. The DOS of CB is much less than any hole band, which results in obvious different distributions between holes and electrons as shown in Fig.1B. For example, at concentration of 3.5E18 /cm$^3$, the holes fill up to 8.5 meV, while the electrons fill up to 63.8 meV. It can be clearly seen that the energy distribution of holes is much less than that of electrons at certain concentration.

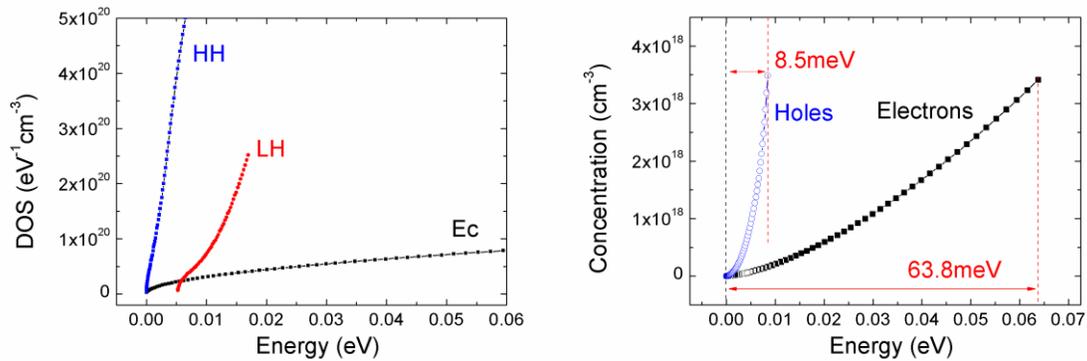

**Figure 1.** The DOS of each band (left), the carrier occupation near the edge of the bands at different concentrations(right).

The calculations of the carrier distribution in momentum space have also been done alone $k_z$ and $k_x$ direction, shown in Fig. 2A and Fig. 2B. It is clear that holes will distribute to higher $k_z$ range than electrons from low concentration and more obvious difference at high concentration, while holes and electrons show the almost same $k_x$ distribution at wide concentration range. From Fig. 2A, the difference of $k_z$ value is 0.0123 Å$^{-1}$ at the concentration of 2E18/cm$^3$, which is much larger than the momentum of a photon (about 2.8E-4 Å$^{-1}$). This means that the huge differences in k value cannot be compensated by a photon, and there is no electron that can match to the hole in the higher $k_z$ range. In order to give a clearer view of the distribution in k space without thermal excitation, a schematic diagram is plotted in Fig. 2C at the concentration of 1.0E18 /cm$^3$, shown in BLUE DOTS, CIRCLES and ARROWS. It is clear that the distribution of the holes is obviously wider than that of the electrons in k$_z$. We call this phenomenon as the **distribution mismatch** between non-equilibrium excess electrons and holes in momentum space. It should be pointed out that this mismatch exists from low concentration, equivalent to from the bottom of the bands. It is well-known that the most radiative recombination occurs near the band bottom, our results indicate that definitely some holes and electrons cannot achieve radiative recombination due to unable to meet the conservation of momentum. in another words, the mismatch caused a loss in the radiative recombination efficiency, higher concentration, more remarkably loss.

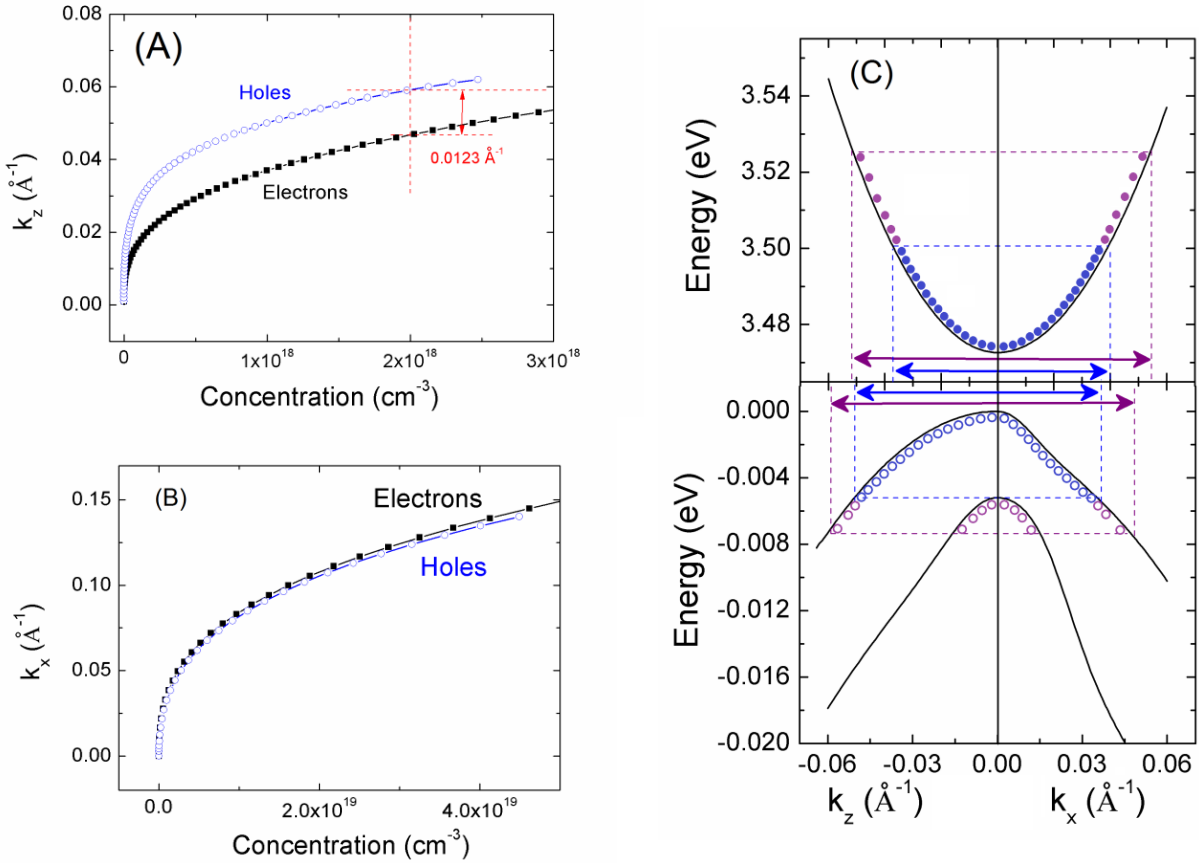

Figure 2. The relationship of carrier concentrations with band filling in k space near the edge of the bands, k$_z$ (A), k$_x$ (B). The schematic diagram of the carrier distribution in CB and VB near the band edge at different temperatures, low temperature in blue markers and higher temperature in purple markers(C).

When it is at a higher temperature, thermal excitation cannot be ignored and both electrons and holes will gain thermal energy to increase distribution range in momentum space. However, because of different band structures between CB and VB, the carrier distribution will become very interesting. There is only one band for electrons, but two bands for holes. The two hole bands, HH and LH, are separated by around 5~10 meV that is equivalent to the thermal energy around 58~115K. This means that the holes in the HH band can be easily excited to the LH band at the temperature higher than 58K. It has been known that the DOS of the LH band is in the same order of magnitude to the HH band. So, holes can distribute in both HH and LH by thermal excitation, the momentum range will not extend as much as those electrons by thermal excited. In the total effect, the momentum increment for holes is smaller than that for electrons with the thermal excitation which reduces the mismatched k value, finally resulting in a larger momentum overlapping range, as shown in PURPLE DOTS, CIRCLES and ARROWS in Fig2C.

In another words, the distribution mismatch between electrons and holes in momentum space can be partially compensated at higher temperatures. This model can explain the temperature-dependence of the internal efficiency droop in GaN-based LEDs, which show a weaker droop effect at higher temperature reported by several groups, including J. Hader (Ref. 20). In Ref. 20, with the increase of the temperature, the maximum IQE condition moves to higher current density (equivalent to higher carrier concentration), and the relative loss of the IQE at the highest current density to the maximum IQE becomes smaller. These show weaker droop effect at higher temperatures, instead of stronger as predicted by normal thinking. From our model discussed above, based on the distribution mismatch between non-equilibrium excess electron and hole in momentum space, it can be easily explained without any assumptions. This is because the holes have to be redistributed to the HH and the LH at higher temperatures, and then the degree of the mismatch in momentum space becomes weaker, finally the IQE loss becomes smaller.

## 3. Optical characterization of high-quality GaN layer

In order to confirm whether the distribution of the nonequilibrium excess carriers is consistent with the results analyzed above, we performed PL measurement at 6 K with different excitation power density. The high-quality GaN layer with no heterostructure was epitaxially grown on a freestanding GaN substrate (GaN/GaN) using metal organic chemical vapor deposition. Normal GaN, grown on a sapphire substrate, shows a dislocation density as high as $10^9/cm^2$. However, our GaN/GaN sample shows very low dislocation density as low as $10^5/cm^2$ examined by high-resolution XRD (Supplementary Fig. S2) and very low doping concentration as low as $10^{15}/cm^3$ measured by Hall measurement. The PL measurements are carried out by a Renishaw micro-PL system. A low-temperature sample stage cooled by a Helium Closed Cycle Cryostat can go down to 6 K, for which the thermal excitation can be ignored. The excitation source is a He-Cd laser with 325 nm and can be focused to a small area about 4 micrometer in diameter. Because the dislocation density of $10^5/cm^2$ is equivalent to an average distance of 30 micrometer between two adjacent dislocations, so it is dislocation free with high possibility inside the focus spot area. The power density of the laser beam is from 75W/cm² to 150kW/cm² in the focus spot area. Thus, the non-equilibrium excess carrier concentration generated in GaN is from 2.5E15/cm³ to 5.0E18/cm³ (Supplementary Fig. S3), i.e., equivalent to ranging from a weak injection to a medium strong injection.

PL process locally generates the same amount of nonequilibrium excess holes in the VB and electrons in CB at the same time. There are several transitions to realize radiative recombination, such as electron-hole pair (excitons) and the transition by the assistance of LO phonons. Although the coupling possibility between carriers and phonons is very low, the relatively large momentum value of the LO can assist the electron-hole to meet the conservation of momentum. As a result, all nonequilibrium excess electrons and holes can contribute to the radiative recombination with the assistance of the LO phonons. Thus, the PL spectrum from the radiative recombination with the assistance of the LO phonons can present the full picture of carrier distributions.

## 4. Results and discussion

The PL spectra of GaN/GaN with different excitation power densities are shown in Fig. 3A. Due to the high crystal quality and low background concentration, each transition is well marked. The detailed band-edge emissions can be clearly observed as marked in Fig. 3B, including free exciton A (FXA), free exciton B (FXB), free exciton C (FXC), donor-bound exciton (DX), and first-order LO replica (1LO).

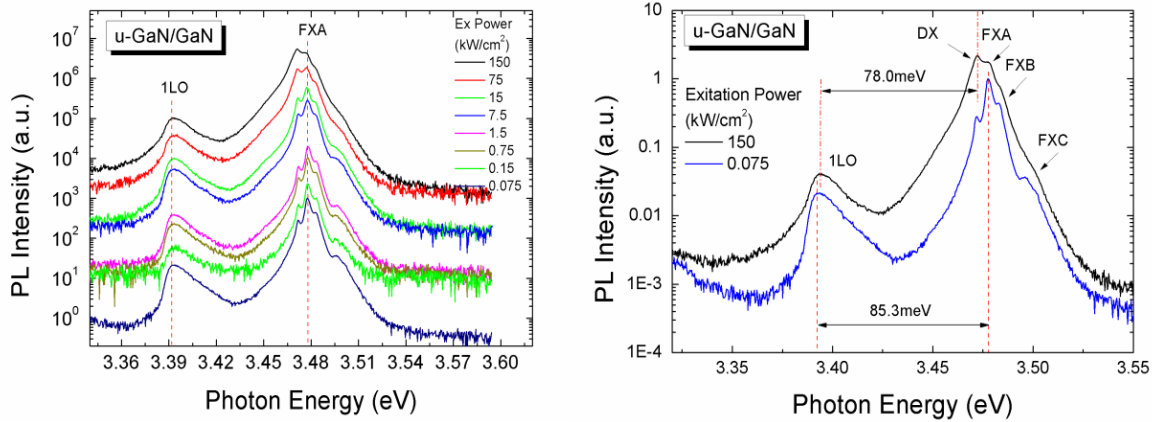

Figure 3. PL spectra of sample GaN/GaN under different excitation intensities at 6K (A). The blue shift of the LO peak position at the lowest and highest excitation levels (B).

The peak positions of all free excitons and DX largely remain unchanged as the electrons occupying at higher energy level can normally relax to the CB bottom within a very short time around 1 ps once there are empty states[21]. This also confirmed no heating effect in the measurement. At higher excitation power density, the FX and DX become broader, which presents more electrons/holes in a wider energy range contributing to the direct radiative recombination. As mentioned above, not all carriers can recombine directly, but can be that with the assistance of the LO. Thus, the peak profile of the 1LO replica can present the distribution of all carriers. In another words, if the peak position of the 1LO moves to a higher energy level, this indicate a wider distribution of the non-equilibrium excess carriers, more than at the band bottom.

Form Fig.3A, it can be seen that the main transition peak changes from FXA to DX with the increase of excitation levels due to different exciton reaction. It then becomes important to determine which transition is the host peak for the 1LO replica. In order to do so, the low-energy edge of the 1LO was compared to the main transition peak in detail, as shown in Supplementary Fig. S4. At the lowest excitation level, the low energy edge of the 1LO peak coincides with that of the FXA. At the highest excitation level, the low energy edge of the 1LO peak coincides with that of the DX. This indicates that the host peak for the 1LO is FXA at the lowest and DX at the highest excitation level, respectively, as shown in Fig.3B. Finally, we found that the 1LO shows a clear blue shift of 7.3 meV (85.3 minus 78.0 meV).

In order to determine which kind of carriers causes the blue shift, we recall Fig. 1B. It is most noticeable that the blue shift value (7.3meV) of the 1LO replica in Fig. 3 coincides with the hole filling level (8.5meV) in Fig. 1B, instead of the electrons. This can be understood by referring to the high $k_z$ range in Fig. 2, only those holes with higher $k_z$ in momentum space do need the assistance of LO phonons to achieve radiative recombination, because their $k_z$ values are larger than all electrons. So, these holes become mismatched holes.

Excitons are well known as the most efficient means for spontaneous emission in normal conditions, but this transition requires the conservation of both energy and momentum, those mismatched holes cannot achieve radiative recombination, and will lose through the non-radiative recombination process with high probability. Thus, the efficiency of the recombination through exciton transition is reduced with the increase of the carrier concentration. On the other hand, all holes and electrons have the chance to realize radiative recombination by the assistance of LO phonons independent on the carrier concentration, although the probability of carriers coupling with phonons is low.

The above conclusion can be supported by the investigation of the light efficiency of the sample. From Fig. 3A, we obtained the integral intensity of each transition from each curve. Based on the excitation power densities, we can obtain the relative emission efficiency for each recombination, shown in Fig. 4, including FX, DX, and 1LO.

From Fig.4, it is clear that the direct recombination, i.e. FX and DX, showed an obvious efficiency droop behavior similar to that in a typical LED[8], which is not unexpected as emission through FX or DX also is typical mechanism in a typical LED. We can also confirm that a more serious efficiency droop can be observed in a GaN layer grown on a sapphire with poorer crystal quality, which the main transition is from DX (Supplementary Fig. S5) and the emission efficiency(black tangle) is also plotted in Fig. 4.

At the same time, the emission efficiency of the 1LO **DOES NOT** show any droop as shown in Fig. 4, although the intensity is low due to low coupling probability. Considering again that the phonons can greatly compensate the momentum mismatch between the holes and electrons, thus is not surprising that the 1LO transition is almost independent from the carrier concentration, even at high excitation power densities.

From our results, it can be seen that improving the quality of the material can help to maintain the emission efficiency. On the other hand, if the momentum mismatch between the holes and electrons can be compensated by some methods, such as by phonons, then the emission efficiency may be less dependent on or even independent of the carrier concentration.

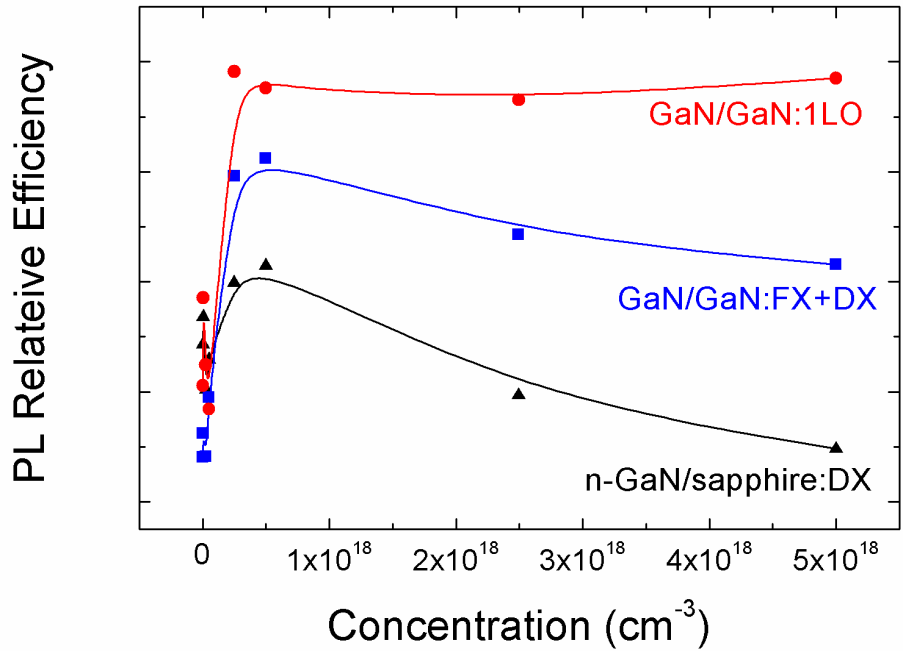

Figure 4. The emission efficiency of 1LO replica and band edge emissions.

From above study, it can be seen that in order to fundamentally eliminate adverse effects of the distribution mismatch between electrons and holes in momentum space, it is necessary and important to investigate new recombination process. A few recent reports have presented some clues, such as that coupling the surface plasmon with InGaN quantum wells can greatly improve the emission performance of the coupled structure[22]. Our group also explored a creative structure to stimulate the radiative recombination between electrons and holes by using surface plasmon[23] that leads to a noticeable improvement.

## Conclusions

We investigate the cause of efficiency droop in GaN material based on GaN intrinsic energy band structure, excess electrons and holes distribution, and conservation of energy/momentum for radiative recombination. From theoretical calculations and vary-excitation PL measurements at low temperature, the distribution mismatch between holes and electrons in momentum space is discovered, and the results confirmed that the distribution of the holes is wider than the electrons.

The distribution mismatch is induced by the energy band parameters of GaN, and it can be partially compensated at higher temperatures. We believe this mismatch is one of contributing factors for the efficiency droop in GaN material. Therefore, for efficient radiative recombination at high carrier concentration, the hole distribution must be modulated and certain new stimulation mechanisms should be explored.

**FUNDING**

This work is supported by National Natural Science Foundation of China (12074182), Collaborative Innovation Center of Solid-State Lighting and Energy-saving Electronics.

**AUTHOR DECLARATIONS**

**Conflict of Interest**

The authors have no conflicts to disclose.

**Author Contributions**

**Peng Chen**: Conceptualization (lead); Funding acquisition (lead); Project administration (lead); Writing - review & editing (lead). **Zili Xie**: Methodology (equal); Project administration (equal). **Xiangqian Xiu:** Resources (equal). **Dunjun Chen**: Methodology (equal). **Bin Liu**: Methodology (equal). **Hong Zhao**: Resources (equal); **Yi Shi**: Validation (equal). **Rong Zhang**: Supervision (equal). **Youdou Zheng**: Supervision (equal).

**DATA AVAILABILITY**

The data that support the findings of this study are available from the corresponding authors upon reasonable request.

# Study on the efficiency droop in high-quality GaN material under high photoexcitation intensity

Peng Chen*, Zili Xie, Xiangqian Xiu, Dunjun Chen, Bin Liu, Hong Zhao, Yi Shi, Rong Zhang, Youdou Zheng[1], and

*Jiangsu Provincial Key Laboratory of Advanced Photonic and Electronic Materials and School of Electronic Science and Engineering, Nanjing University, Jiangsu, Nanjing, 210093, China.*
*Authors to whom correspondence should be addressed: pchen@nju.edu.cn

## Supplementary Information:

### I. GaN band structure calculation

Solving the 6×6 K·P Hamiltonian yielded the CB and VB of GaN, as shown in Fig.1. As previously reported, the energy bands are asymmetric along the $k_z$ and $k_x$ directions.

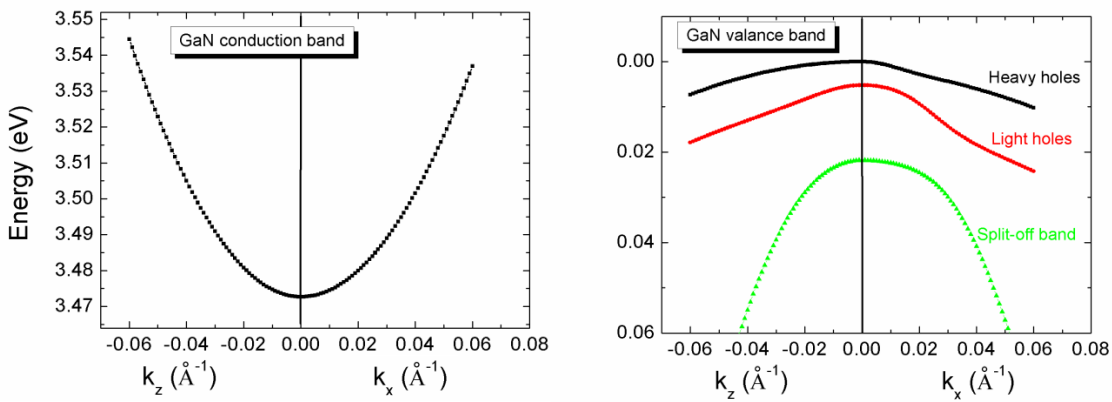

Fig.S1. GaN band structure near the CB bottom (left) and the VB top(right).

### II. High-quality GaN layer grown on the GaN substrate

A high-quality GaN layer was grown on a freestanding GaN substrate. High-resolution XRD results are shown in Fig. 2. The full width at half maximum (FWHM) of the (0002) diffraction peak is more than 400 arcseconds for a normal GaN layer grown on a sapphire substrate, while the FWHM is only 72 arcseconds for the high-quality GaN layer grown on the GaN substrate. The GaN/GaN sample shows extremely low dislocation density — as low as $10^5/cm^2$, as calculated from the high-resolution XRD result.

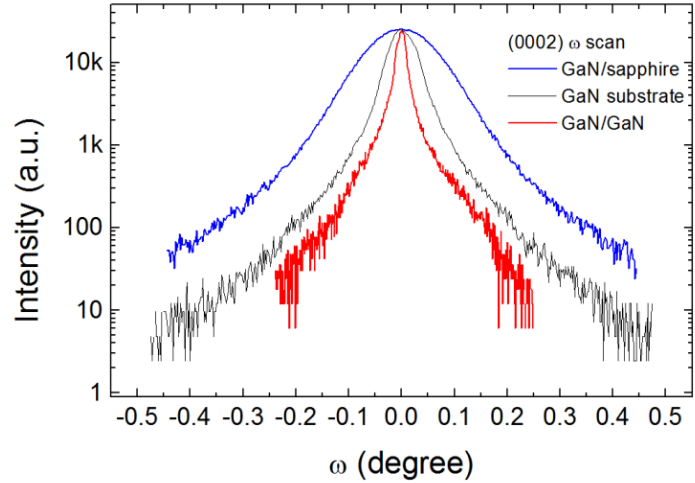

Fig.S2. Comparison of crystal quality between GaN/sapphire and GaN/GaN by high-resolution XRD measurement.

## III. Calculation of the nonequilibrium excess carrier concentration generated in the PL experiments

The PL measurements were carried out by a Renishaw micro-PL system, the diameter of the focus spot was around 4 μm. The excitation source was a He-Cd laser with a wavelength of 325 nm. The samples were put into a low-temperature stage cooled by a helium closed-cycle cryostat, and the temperature could reach as low as 6 K. During the measurements, the highest power of the laser beam on the sample surface was about 20 mW, equivalent to the power density of 150 kW/cm$^2$ in the focus area. The different excitation power densities were obtained by using different neutral density filters.

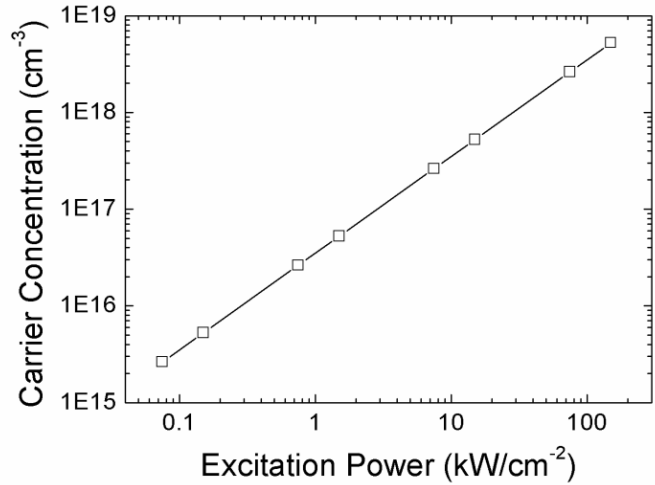

Fig. S3. The relationship between the photo-generated nonequilibrium extra carrier concentration and the excitation power density.

Fig. 3 correlates the laser power density and GaN absorption behavior, plotting the carrier concentration generated in GaN by different laser power densities. The carrier concentration ranges from 2.5E15/cm$^3$ to 5.0E18/cm$^3$, i.e., equivalent to ranging from a weak injection to a medium-strong injection.

IV. **Host peaks of the 1LO**

To determine the host peaks of the 1LO, we made a detailed comparison of the low energy edge of the 1LO and the profile of the band edge emissions, as shown in Fig. 4. At the lowest excitation level, the low energy edge of the LO peak coincides with that of the FXA. At the highest excitation level, the low energy edge of the LO peak coincides with that of the DX. This indicates that the host peak for the 1LO is the FXA and the DX at the lowest and highest excitation levels, respectively, as shown in Fig. 4.

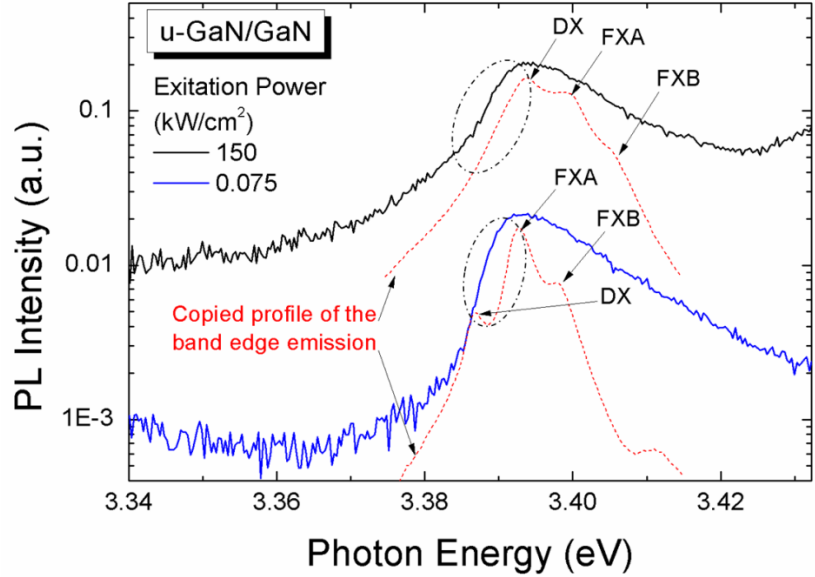

Fig. S4. Comparison of the low energy edge of the 1LO and the profile of the band edge emissions.

## V.  PL results of a normal GaN layer grown on sapphire substrate

For comparison, a normal quality GaN layer was grown on a c-plane sapphire substrate (GaN/sapphire) with slight Si doping. The GaN/sapphire shows normal quality as most paper reported, the dislocation density is in $10^8/cm^2$ range and electron concentration is in $1E17/cm^3$ range.

Fig.5 shows the PL spectra of a GaN/sapphire. Because this sample's crystal quality was relatively lower and slightly doped, each transition became broader. As shown, the main peak of the spectra was always DX, which means that DX was always the host peak of LO in this sample. The 1LO shows a clear blue shift of about 5.3 meV.

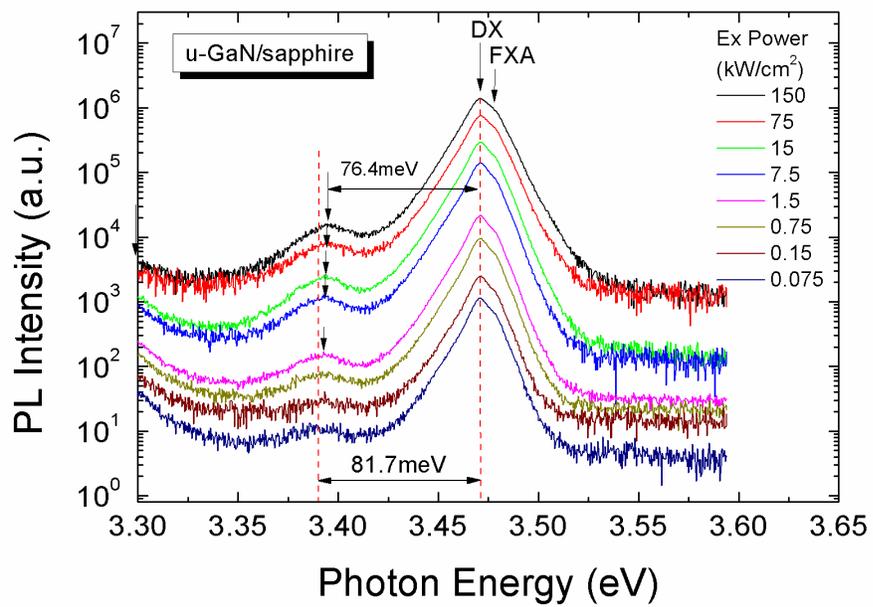

Fig. S5. The PL spectra of the GaN layer grown on a sapphire substrate.